\input psfig
\documentstyle[11pt]{article}
\def\be{\begin{equation}}
\def\ee{\end{equation}}
\def\la{\label}
\def\bea{\begin{eqnarray}}
\def\eea{\end{eqnarray}}
\def\non{\nonumber} 
\def\ci{\cite}
\def\la{\label}
\def\fr{\frac}
\def\pp{\partial}

\def\cOm{\tilde{\Omega}}
\def\bib{\bibitem}
\def\le{\left}
\def\ri{\right}

\def\gm{\gamma}
\def\al{\alpha}

\def\Lm{\Lambda}
\def\lm{\lambda}

\begin{document}

\begin{flushright}
astro-ph/9911079 \\
IFUNAM-FT-9909 \\

\end{flushright}
\vspace{15mm}

\begin{center}
{\Large \bf  Model Independent Accelerating Universe  
and the Cosmological Coincidence Problem
} \\
\end{center}
\vspace*{0.7cm}

\begin{center}

{\bf A. de la Macorra\footnote{e-mail: macorra@fenix.ifisicacu.unam.mx} }
\end{center}

\vspace*{0.1cm}
\begin{center}
\begin{tabular}{c}
{\small $$Instituto de F\'{\i}sica, UNAM}\\
{\small Apdo. Postal 20-364}\\
{\small 01000  M\'exico D.F., M\'exico}\\
 \end{tabular}
\end{center}

\vspace{1 cm}

\begin{center}
{\bf ABSTRACT}
\end{center}

{\small   

We show that  the evolution of the quintessence energy density    $\Omega_Q$ is model independent in an accelerating universe. The  accelerating behaviour has lasted at most 0.5 e-folds of expansion assuming a  present day value of  $\Omega_Q=2/3$.  The generic  evolution differs from the exact solution
by a  percentage given by $r=2\gm_Q$, with $\gm_Q=(\rho_Q+p_Q)/\rho_Q$. For a small $\gm_Q$ the evolution remains a good approximation for a long period. Nucleosynthesis bounds on $\Omega_Q$ suggest that the model independent solution is valid for at least 12 e-folds of expansion which includes then the scale of  radiation and matter equality. We can, therefore, establish model independent conditions on the cosmological parameters
at some factor  scale  $a_i  \ll a_o$.  Finally,  we discuss the  relevance of this result to  the cosmological coincidence  problem. 

}

\noindent

\rule[.1in]{14.5cm}{.002in}

\thispagestyle{empty}

\setcounter{page}{0}
\vfill\eject

Recent observations show that the universe has entered an accelerating expansion regime \ci{Riess}. If these observations are confirmed it would be the first experimental evidence for  an energy density different  from the radiation or matter and with negative pressure, i.e. for a "cosmological  constant". The cosmological constant can be given by   constant energy density
with equation of state $p_\Lambda=-\rho_\Lambda$, where $p_\Lm$ is the pressure and $\rho_\Lm$ the energy density, or it could be parameterised in terms of a slow varying scalar field Q, quintessence, with a potential $V(Q)$ and  a time varying equation of state $p_Q=(\gm_Q-1) \rho_Q$, with   $p_Q=\fr{1}{2}\dot Q^2-V(Q)$ and $\rho_Q=\fr{1}{2}\dot Q^2+V(Q)$. The equation of state has $0 \leq \gm_Q \leq 2$, it is  model dependent and it may vary with time.  For an accelerating universe we have the condition $ (\gm_Q-1)\Omega_Q < -1/3$, where $\Omega_Q$ is the ratio between the energy density of $Q$ and the critical energy density.

An important
question is why the energy density of the cosmological constant is of the same order of magnitude as matter even though the expansion rate of both energies is quite different today.  This is the coincidence problem \ci{tracker}. A possible answer is by setting the initial conditions such that the Q field start to dominate only recently. However, this solution introduces a fine tuning problem on the initial conditions. To avoid such a fine tuning problem {\it tracker} solutions have been proposed \ci{tracker}. In this  models the energy density of the scalar field Q redshifts mimics the dominant energy density (radiation) at the beginning with $\gm_Q > 1$ and when matter dominates  the equation of state for $Q$ changes to $\gm_Q < 1$. With this kind of fields one avoids the initial fine tuning problem but they have difficulties to explain (without any fine tuning of the potential) the central values of $\Omega_Q=2/3 \pm 0.05, \, \gm_Q \leq 0.35 \pm 0.07$  \ci{fits} since they have   $\gm_Q > 0.3$ \ci{tracker}. However, they remain a very interesting possibility. Other models with more the one term in the potential $V=V_1+V_2$ where one term dominates during radiation and scales as radiation or matter while the other dominates at a recent time giving the acceleration of the universe have been studied \ci{v+v}. In particular if during the radiation dominated epoch on has a potential $V_1=a\,Q^4$, it will redshift as radiation and the ratio $\rho_r/\rho_Q$ remains constant. On the other for  $V_1=a\,Q^2$ the ratio $\rho_m/\rho_Q$ is constant. In either case the initial value can be set to $\rho_r/\rho_Q=O(1)$ or $ \rho_r/\rho_Q=O(1)$ avoiding an initial fine tuning problem but not the coincidence problem. Finally, we can invoke the anthropic principle to try to explain the coincidence problem where the likelihood of our present day cosmological parameters are determined  \ci{anthro}.

Different models have been proposed to give an accelerating universe and a general analysis can be found in \ci{general}, \ci{mioscalar}. We have 
the exponential potential, $V \sim e^{-\lambda Q}$ with $\lm < 3$ \ci{vexp},\ci{liddle}, the inverse power potentials $V \sim 1/Q^{n}, \; n>0$ \ci{1/q},\ci{tracker},  $V\sim e^{1/Q}$ \ci{exp1/q} or $V\sim (c + (Q-Q_0)^n)^a e^{-\lm Q}$ \ci{vp} and mixtures of any of these potentials. All these potentials have a finite $\lambda=-V_Q/V < O(1) $, with $V_Q \equiv \pp V/ \pp Q$,  when  $V$ approaches its minimum and this is indeed required   for any model to give an accelerating universe since in this limit the potential energy dominates the energy density of Q. Some of these potentials may arise from non-perturbative effect or form string theory \ci{miomod}, \ci{mod}.

In this work we address the reverse coincidence problem. This is,  we start by imposing  present days values of  $\Omega_Q, \Omega_m$ and $\gm_Q$ and we go backwards in time. We will show that the evolution of $\Omega_Q$ while the universe accelerates is model independent. We can therefore establish generic  initial conditions for $\Omega_Q$ at   the beginning of the accelerating epoch. Furthermore,  as long as $\gm_Q$ is small (say $\gm_Q < 0.4$)  the evolution of $\Omega_Q$ remains model independent (within an error of $r=2\gm_Q=20\%$) and this allows to set the initial conditions at a much smaller scale factor $a_i$ compared to today's value $a_o$, e.g. at radiation and matter equality $a_i/a_o \simeq 10^{-3}$  or at nucleosynthesis $a_i/a_o\simeq 10^{-10}$. The range of validity is model and initial  conditions dependent,  however, in most models the universe have a scaling regime with $\Omega_Q \ll 1,\, \gm_Q \ll 1$ before the acceleration epoch leading to a large number of e-folds of model independent expansion. This regime seems to be necessary due to the condition $\Omega_Q < 0.1$ at nucleosynthesis \ci{Freese}. Here, we do not solve the coincidence problem but we contribute to establish model independent behaviour of the universe at the accelerating era and some e-folds before that, say $a_{m.i.}$. The last step would be to mach those initial conditions at $a_{m.i.}$ with a particular model coming from earlier times.

In a  spatially flat Friedmann--Robertson--Walker (FRW) Universe, the Hubble
parameter is given by $H^2=\fr{1}{3}\rho=\fr{1}{3}(\rho_b+\rho_Q)$ where the   barotropic fluid is described by an energy density $\rho_{b}$ and a pression $p_{b}$ with a  standard equation of state  $p_{b}=(\gm_b-1) \rho_{b}$, with $\gm_{b}=1$ for matter and $\gm_{b}=4/3$ for radiation. Since we will study the evolution starting from today, we will take $\gm_b=0$, however we will leave the $\gm_b$ dependence in the equations. The interaction between the barotropic fluid and Q is gravitational only (we take $8\pi G=1$). In terms of the of the variables  $x \equiv  {\dot Q} / \sqrt {6} H$, $y \equiv  \sqrt{ V / 3} H$ \ci{liddle} the cosmological evolution of the universe is given by
\bea
x_N&=& -3 x + \sqrt {3 \over 2} \lambda\,  y^2 + {3 \over 2} x [2x^2 + \gm_b (1 - x^2 - y^2)]  \non \\
y_N&=& - \sqrt {3 \over 2} \lambda \, x\, y + {3 \over 2} y [2x^2 + \gm_b (1 - x^2 - y^2)]
 \la{cosmo1} \\
H_N&=& -{3 \over 2} H [\gm_b (1-x^2-y^2) + 2x^2] \non
\eea
where $N$ is the logarithm of the scale factor $a$, $N \equiv ln (a)$, $f_N\equiv df/dN$ for $f=x,y,H$ and $\lambda (N) \equiv - V_Q/ V$. Notice that all model dependence in eqs.(\ref{cosmo1}) is through the quantities $\lm (N)$ and the constant parameter $\gm_{b}$ and generic solutions can be obtained \ci{general},\ci{mioscalar}. The flatness condition gives $1=\Omega _Q + \Omega_b$ and from eqs.(\ref{cosmo1}) the evolution of $\Omega_Q=x^2+y^2$ and $ \gm_Q=2x^2/(x^2+y^2) $ is
 \bea
   (\Omega_Q)_N&=& 3 (\gm_b-\gm_Q) \Omega_Q (1- \Omega_Q)
\non\\
(\gm_Q)_N &=& \gm_Q (2-\gm_Q)(-3+ \sqrt{\fr{3}{2}} \fr{\lm}{x} \Omega_Q )
\la{dOm}
\eea
An  accelerating expansion universe requires $\rho+3p=\rho_b (3\gm_b-2)+\rho_Q (3\gm_Q-2) <0$ which gives the constraint
\be
(\gm_Q -1)\Omega_Q < -1/3
\la{wOm}\ee
where we have taken the barotropic fluid as matter, i.e. $\gm_b=1$.   Equivalently, we can write  condition (\ref{wOm}) in terms of $x$ as $x^2 < (\Omega_Q-1/3)/2$ and at present day this sets an upper value of $x^2\leq 1/6$ and $\gm_Q < 1/2$ for today's central value of $\Omega_{oQ}=2/3$ (from now on the subscript $o$ refers to present day quantities). If we take  the preferred  value $ \gm_{oQ} \leq 0.35$  the condition on $x$ is  $x^2_o= \fr{1}{2}\Omega_{oQ} \;\gm_{oQ} \leq 0.11 $, which is slightly smaller then the previously obtained one,  and $y^2_o=\Omega_{oQ}(1-\gm_{oQ}/2) \geq 0.55$, i.e. we have  $y^2_o \geq 5\,x^2_o$. The solution to eq.(\ref{dOm}) is only model dependent through
the quantity $\lm(N)$.  Solving eq.(\ref{dOm}) for $\Omega_Q$ we obtain 
\be
\Omega_Q(N)=\fr{\Omega_{oQ} e^{-3\gm_b(N_o-N)+3\int\gm_Q dN}}{1-\Omega_{oQ} + \Omega_{oQ}e^{-3\gm_b(N_o-N)+3\int \gm_Q dN}}
\la{Om}
\ee
where $\Omega_{Q}(N_o)=\Omega_{oQ}$ and we have $N < N_0$ for a time $t < t_0$. In the limit of $\Omega_{oQ} \ll 1$ and $\gm_Q \simeq cte$ eq.(\ref{Om}) reduces to $\Omega_Q \simeq e^{3N(\gm_b-\gm_Q)}$. However we are not interested in this region since the initial condition has $\Omega_{oQ} \simeq 2/3$. During the
accelerating period we have $x^2 < y^2$ and $\gm_Q < \gm_b$ and we can expand eq.(\ref{Om}) as a function of $R \equiv  3\int\gm_Q dN$. Eq.(\ref{Om}) becomes 
\be
\Omega_{Q}=\cOm_Q \le(1+\fr{1-\Omega_{oQ}}{1-\Omega_{oQ}+\Omega_{oQ}e^{-3\gm_b(N_o-N)}} R + O(R^2)\ri)
\la{OmR}
\ee
with
\be
\cOm _Q=\fr{\Omega_{oQ}e^{-3\gm_b(N_o-N)}}{1-\Omega_{oQ} +
\Omega_{oQ} e^{-3\gm_b(N_o-N)}}.
\la{OmI}
\ee
Notice that as long as $R$ is constant or $R\ll 1$, eq.(\ref{OmR}) is model independent and it is certainly valid  during the acceleration period of the universe. We see, therefore, that the accelerating expansion of the universe is model independent and  the energy density of quintessence  evolves as $\cOm_Q$.  Furthermore, $\cOm_Q$ remains a good approximation to eq.(\ref{Om})  as long as $\gm_Q$  is small. We can establish a relationship between the amount of discrepancy that we wish to accept between  $\Omega_Q$ and $\cOm_Q$. For a $r$ percentage of difference, i.e. $\Omega_Q/\cOm_Q=1+r$ we have $r=\fr{1-\Omega_{oQ}}{1-\Omega_{oQ}+\Omega_{oQ}e^{-3\gm_b(N_o-N)}} R \simeq R$ (at large $N_0-N$). Since $(\gm_Q)_N$ is proportional to $\gm_Q$ it varies rapidly when $\gm_Q$ is not small  as can be seen from eq.(\ref{dOm}). $R$ becomes large when $\gm_{QN}\simeq -6\gm_Q$ (as a first approximation) and  we find $R=3\int \gm_Q dN= (\Delta \gm_Q)/2$. If we allow an $r=20\%$ discrepancy between the exact and the model independent $\tilde\Omega_Q$, this will happen at   $\Delta \gm_Q =2 r= 0.4$. Since before the accelerating epoch one has $x \ll y$
and $\gm_Q \ll 1$, the model independent energy density  stops being a good approximation only until $\gm_Q\simeq 2r$. When will this happen  is model and initial condition depended, however, for   a large number of cases $N_0-N_i$ can be quite large ($N_i$ is the beginning of the model independent evolution of Q). In fact, if ,for example, we set initial conditions such  that  $x < 2 y$, i.e. $\gm_Q < 0.4$, at $N_i$ (regardless of its value) 
the evolution of $\Omega_Q$ is model independent (within a maximum of 20\% discrepancy). 
 
\begin{figure}
\psfig{file=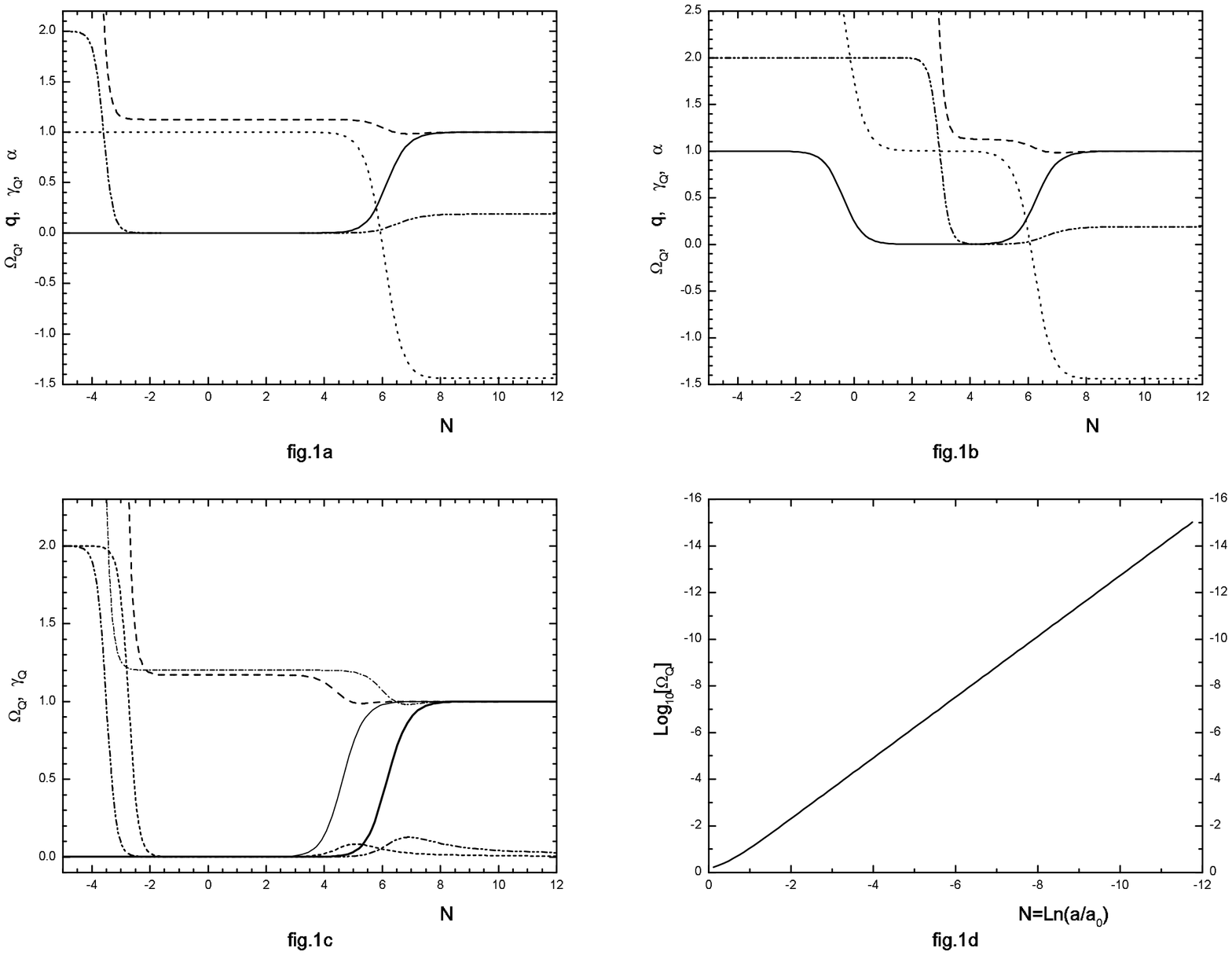,width=18cm,height=18cm}
\caption{\footnotesize{ In fig.1a and 1b we show the behaviour of $\Omega_Q$ (solid line), $\gm_Q$ (dash-dot), $\alpha=\rho+3p/(\rho_b+3p_b)$ (short dashed line) and $\Omega_Q/\cOm_Q$ (long dashed line) for an exponential  potential with 
$\lm=3/4$ and for two different initial conditions at $N=0$, $x=0, y=0.001$ and $x=0.5, y=0.001$ for graph 1a and 1b respectively. In fig.1c we show the behaviour of $\Omega_Q$, $\gm_Q$ and the ratio $\Omega_Q/\cOm_Q$ for a potential $V \simeq e^{-1/Q}$ (thin solid line, short dashed and long dashed lines respectively) and for a potential
$V\simeq 1/Q$ (thick solid line, dashed-dot-dot and dashed-dot line respectively) with the same initial conditions $x=0, y=0.001$ at $N=0$.  Notice that in all cases the model independent solution agrees quite well with the exact solution, for a large number of e-folds and  the discrepancy  becomes significant when $\gm_Q$ starts to increase. Finally, in  fig.1d we plot the $\cOm_Q$ vs N. 
 }}
 \end{figure}

In fig.1a and 1b we show the behaviour of the exact $\Omega_Q$ by solving the dynamically eqs.(\ref{cosmo1}) numerically compared to $ \cOm_{Q}$ for the same exponential potential $V=V_0 e^{-3Q/4}$ but with different initial conditions  $x=0,\,y=0.001$ and $x=0.5,\, y=0.001$ respectively at  $N=0\neq N_o$. 
To appreciate more the difference between these two quantities we plot 
$\Omega_{Q}/\cOm_{Q}$ and we include in the graph
the acceleration parameter $\al=\rho+3p/(\rho_b+3p_b)$ \ci{mioscalar} and $\gm_Q $. We observe from the graphs that the concordance between $\cOm_Q$ and $\Omega_{Q}$ is remarkable as long as the universe expands in an accelerating way ($\al <0$) and $\gm_Q $ remains small.
We would like to mention that running the models down from present day the solution is very sensitive to the initial conditions at $N_o$. This is because we have a late time attractor and models with small variations at a late time may have come from  quite different values at $\Delta N$ large. We also show 
in fig.1c two different models ($V \simeq 1/Q$ and $V \simeq e^{-1/Q}$) with initial conditions $x=0,\,y=0.001$ at $N=0$.  

Form fig.1  we see that as long as $x,\, \gm_Q$ are small the agreement  between $\Omega_Q $ and $\cOm_Q$ is very good. 
In order to avoid a rapid increase in $\Omega_Q$ (going backwards in time), which would contradict the nucleosynthesis bound ($\Omega_Q < 0.1$), we require precisely this kind of behaviour.  So, we expect our $\cOm_Q$ to be a good approximation for a long period of expansion that could go as far as matter and radiation equality or nucleosynthesis. Furthermore, if we run the models backwards the "early" time attractor has $\Omega_Q=1$ (with $x=1,\,y=0$) and the scale at which the transition between $\gm_Q > 1$ and $\gm_Q < 1$ takes place  is a symmetric point in $\Omega_Q$ as a function of $N$. Therefore if $\Omega_Q \ll 1$ for $\Delta N=2M$ then our generic solution will be valid for at least $M$ e-folds of expansion.  If we assume that the quintessence energy density has been increasing since nucleosynthesis (i.e. no extra complications) then $\cOm_Q$ is valid for at least 12 e-folds, i.e. it includes matter and radiation equality scale. 

We can invert  eq.(\ref{Om}) to get the number e-folds of expansion $\Delta N= N_o-N_i$ (see fig.2b) as a function of  $\Omega_{i Q}(N_i)/\Omega_{o Q}(N_o)$,
\be
\Delta N= \fr{1}{3}Log\le[\fr{\Omega_{oQ}}{\Omega_{iQ}}\le(\fr{1-\Omega_{iQ}}{1-\Omega_{Q}}\ri)-Log[1+R]\ri] 
\la{N}
\ee
We can therefore determine the amount of quintessence  $\Omega_{ Q }$ at different earlier times. For example, the number of e-folds of accelerating expansion is at most $\Delta N=0.46$
since $\Omega_Q$ must be greater or equal than $1/3$ (cf. eq.(\ref{wOm})). 
During structure formation with $a_i/a_o \simeq 1/2.2$ we have $\Omega_Q =0.15$, at matter and radiation equality (useful for tracker models) $a_i/a_o \simeq 10^{-3}$ we get $\Omega_Q=2\times 10^{-9}$ and  at nucleosynthesis
with $a_i/a_o \simeq 10^{-10}$ we have $\Omega_Q=2\times 10^{-23}$. With these results we can establish the initial conditions from which the model independent 
expansion is valid and we only need to concentrate on the behaviour at earlier times.

To conclude, we have seen that the expansion of the accelerating universe is model independent and it is given by $\cOm_Q$.  As long as $\gm_Q$ is small the evolution of $\cOm_Q$ remains a good approximation to the exact solution and the percentage difference is given by $r=2\gm_Q$. Due to the nucleosynthesis bound ($\Omega_Q <0.1$), we  do not expect (going backwards in time from today's value) $\Omega_Q$ to have a decrease and then a rapid increase and therefore   our solution should be valid for at least  12 e-folds of expansion. We can then impose model independent condition at much smaller scales then our present day one, like at radiation and matter equality. This result  contributes to solve the reverse coincidence problem since we can trace back  the problem of initial conditions to a scale  in which the universe expanded  in a non-accelerating way and possibly as matter or radiation.

  A.M. research was supported in part by CONACYT  project 32415-E  and by DGAPA, UNAM,  project IN-103997.

\thebibliography{}
\footnotesize{

\bibitem{Riess} {A.G. Riess {\it et al.}, Astron. J. 116 (1998) 1009; S.
Perlmutter {\it et al}, ApJ 517 (1999) 565; P.M. Garnavich {\it et al}, Ap.J 509 (1998)
74.}

\bib{tracker} I. Zlatev, L. Wand and P.J. Steinhardt, Phys. Rev. Lett.82 (1999) 8960;  Phys. Rev. D59 (1999)123504

\bib{fits} {G. Efstathiou, S.L. Bridle, A.N. Lasenby, M.P. Hobson and R.S. Ellis, MNRAS 303L (1999) 47; I. Zlatev, L. Wand and P.J. Steinhardt,  astro-ph/9901388; M. Roos and S.M.
Harun-or Rashid, astro-ph/9901234}

\bib{anthro} J. Garriga, M. Livio, and A. Vilenkin, astro-ph/9906210

\bib{general} A.R. Liddle and R.J. Scherrer,Phys. Rev. D59,  (1999)023509;

\bib{mioscalar} A. de la Macorra and G. Piccinelli hep-ph/9909459

\bibitem{Freese} {K. Freese, F.C. Adams, J.A. Frieman and E. Mottola, Nucl. Phys. B 287
(1987) 797; M. Birkel and S. Sarkar, Astropart. Phys. 6 (1997) 197.}

\bib{vexp} C. Wetterich, Nucl. Phys. B302 (1998) 668 
 P. Ferreira, M. Joyce, Phys. Rev. D 58 (1998) 503; E.J. Copeland, A. Liddle and D. Wands, Ann. N.Y. Acad. Sci. 688 (1993) 647.  

\bibitem{1/q} {P.J.E. Peebles and B. Ratra, ApJ 325 (1988) L17;B. Ratra and P.J.E. Peebles, Phys. Rev. D37 (1988) 3406}

\bib{exp1/q} P. Brax and J. Martin, astro-ph/9905040

\bib{vp}  A. Albrecht and  C. Skordis, astro-ph/9908085, V. Sahni and L.Wang, astro-ph/9910087

\bib{v+v} T. Barreiro E.J. Copeland and N.J. Nunes, astro-ph/9910214; I. Zlatev and P.J. Steinhardt, Phys. Lett.B459 (1999) 570.

\bib{liddle} E.J. Copeland, A. Liddle and D. Wands, Phys. Rev. D57 (1998) 4686

\bib{miomod} A. de la Macorra hep-ph/9910330
\bib{mod} M.C. Bento and O. Bertolami, gr-qc/9905075, O. Bertolami and R. Schiappa, Class.Quant.Grav.16 (1999) 2545

}

\end{document}